\documentclass[12pt]{emulateapj}

\usepackage{natbib}

\newcommand{\teff}{${T}_{\mathrm{eff}}$}
\newcommand{\logg}{$\log{g}$}
\newcommand{\msun}{$M_{\odot}$}
\newcommand{\rsun}{$R_{\odot}$}

\newcommand{\omc}{$(O-C)$}

\shorttitle{Rapid Orbital Decay in a 12.75-Min WD+WD Binary}
\shortauthors{Hermes et al.}

\begin{document}

\title{RAPID ORBITAL DECAY IN THE 12.75-MINUTE BINARY WHITE DWARF J0651+2844}
\author{J. J. Hermes\altaffilmark{1,2}, Mukremin Kilic\altaffilmark{3}, Warren R. Brown\altaffilmark{4}, D. E. Winget\altaffilmark{1,2}, Carlos Allende Prieto\altaffilmark{5,6}, \\ A. Gianninas\altaffilmark{3}, Anjum S. Mukadam\altaffilmark{7,8}, Antonio Cabrera-Lavers\altaffilmark{5,6,9}, and Scott J. Kenyon\altaffilmark{3} }

\altaffiltext{1}{Department of Astronomy, University of Texas at Austin, Austin, TX 78712, USA}
\altaffiltext{2}{McDonald Observatory, Fort Davis, TX 79734, USA}
\altaffiltext{3}{Homer L. Dodge Department of Physics and Astronomy, University of Oklahoma, 440 W. Brooks St., Norman, OK 73019, USA}
\altaffiltext{4}{Smithsonian Astrophysical Observatory, 60 Garden St, Cambridge, MA 02138, USA}
\altaffiltext{5}{Instituto de Astrofisica de Canarias, E-38205 La Laguna, Tenerife, Spain}
\altaffiltext{6}{Departamento de Astrofisica, Universidad de La Laguna, E-38206 La Laguna, Tenerife, Spain}
\altaffiltext{7}{Department of Astronomy, University of Washington, Seattle, WA 98195, USA}
\altaffiltext{8}{Apache Point Observatory, 2001 Apache Point Road, Sunspot, NM 88349, USA}
\altaffiltext{9}{GTC Project, 38205 La Laguna, Tenerife, Spain}

\email{jjhermes@astro.as.utexas.edu}


\begin{abstract}

We report the detection of orbital decay in the 12.75-min, detached binary white dwarf (WD) SDSS J065133.338+284423.37 (hereafter J0651). Our photometric observations over a 13-month baseline constrain the orbital period to $765.206543(55)$ s and indicate the orbit is decreasing at a rate of $(-9.8\pm2.8) \times 10^{-12}$ s s$^{-1}$ (or $-0.31\pm0.09$ ms yr$^{-1}$). We revise the system parameters based on our new photometric and spectroscopic observations: J0651 contains two WDs with $M_1 = 0.26\pm0.04$ \msun\ and $M_2 = 0.50\pm0.04$ \msun. General relativity predicts orbital decay due to gravitational wave radiation of $(-8.2\pm1.7) \times 10^{-12}$ s s$^{-1}$ (or $-0.26\pm0.05$ ms yr$^{-1}$). Our observed rate of orbital decay is consistent with this expectation. J0651 is currently the second-loudest gravitational wave source known in the milli-Hertz range and the loudest non-interacting binary, which makes it an excellent verification source for future missions aimed at directly detecting gravitational waves. Our work establishes the feasibility of monitoring this system's orbital period decay at optical wavelengths.

\end{abstract}



\section{Introduction}

The 12.75-minute orbital period detached binary WD system J0651 was discovered by \citet{BrownJ0651} as part of the extremely low mass (ELM, $\leq 0.25$ \msun) WD Survey, a targeted spectroscopic search for ELM WDs. While that survey has yielded some two dozen merger systems, with orbital periods of tens of minutes to hours \citep{BrownELMi,KilicELMii,BrownELMiii,KilicELMiv}, none are as compact as J0651.

In addition to a large radial-velocity amplitude, this double degenerate system is oriented in such a way that it yields a wealth of photometric information: eclipses of each star by the other, ellipsoidal variations and Doppler boosting. While photometric observations engender an accurate way to measure the orbital and system parameters, they also provide multiple clocks with which to monitor the orbital evolution of the system.

The orbital decay of compact binary systems is currently the best method to detect the influence (and existence) of gravitational waves, and few known systems are radiating as strongly or decaying as rapidly as J0651. There are presently just five binaries known with orbital periods less than 15 minutes, and the other four are interacting: three are the AM CVn systems HM Cnc, V407 Vul, and ES Cet, and the other is the low-mass X-ray binary 4U 1820-30 \citep{Israel99,Haberl95,Steeghs06,Warner02,Stella87}. After the 5.4-minute HM Cnc \citep{Israel02,Roelofs10}, J0651 is the second-loudest gravitational wave source known in the milli-Hertz frequency range \citep{Amaro12}. J0651 is thus the shortest-period detached compact binary known and the cleanest system to observe at optical wavelengths for orbital decay due to gravitational wave radiation.

In this Letter we present follow-up photometric and spectroscopic observations of J0651, refine orbital and system parameters, and report the detection of rapid orbital decay in the system. We discuss the orbital period change in the context of expectations from general relativity, as well as deviations expected due to tidal interactions. Section 2 describes our observations, and Sections 3 and 4 present the refined system parameters and the orbital period decay, respectively.



\section{Observations and Reductions}
\label{sec:photobs}

\subsection{Photometric Observations}

Our discovery observations of J0651 ($g_0=19.1$ mag) were described by \citet{BrownJ0651} and included some 12.7 hr of photometry from the McDonald Observatory (McD) 2.1 m Otto Struve Telescope using the Argos frame-transfer camera \citep{Nather04}. In the subsequent year, we have obtained an additional 195.4 hr of photometry using four different facilities.

\begin{figure}[t]
\centering{\includegraphics[width=0.45\textwidth]{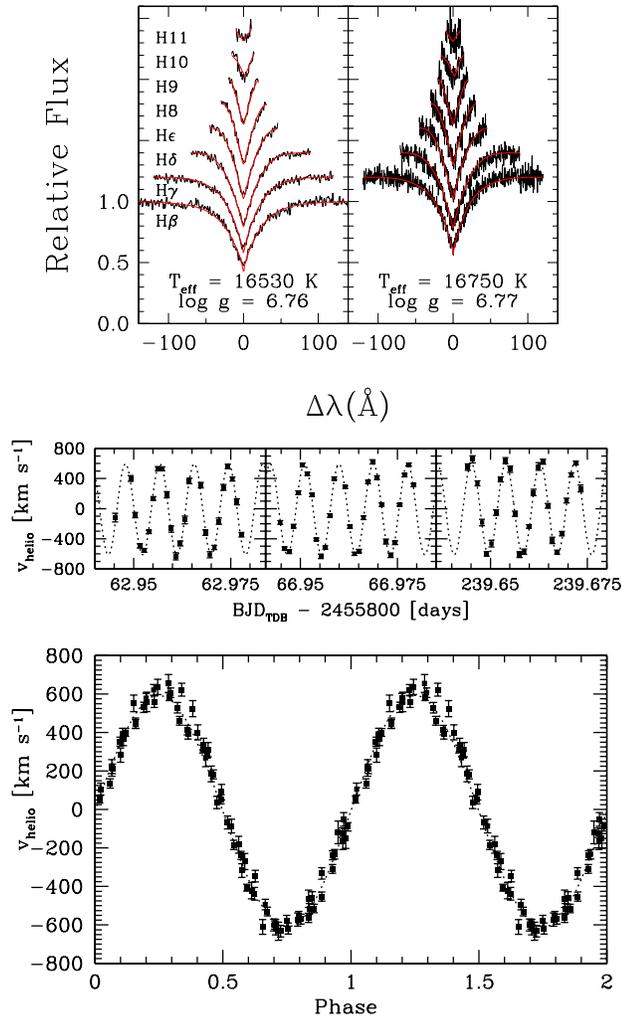}}
\caption{Spectroscopic observations of J0651. The top left panel shows the summed spectra from our October 2011 and April 2012 observations, with model fits to the H$\beta$ to H11 lines, which we use to derive the primary parameters in Section~\ref{sec:param}. The top right panel shows fits to the lower S/N, summed spectra from March 2011 from \citet{BrownJ0651}. The middle panel shows our new radial velocity observations of J0651 from three new epochs, and the bottom panel shows those data phased to the orbital period. \label{fig:J0651newrv}}
\end{figure}

The majority of our continuing photometric observations were carried out in an identical manner to the discovery observations, with the Argos instrument through a 1 mm \textsl{BG40} filter to reduce sky noise. We obtained $5-30$ s exposures of the target, with a typical exposure time of 10 s, depending on the observing conditions. Additionally, we obtained nearly 3 hr of data in December 2011 using the Agile instrument \citep{Mukadam11} mounted on the 3.5 m telescope at Apache Point Observatory (APO), using a 1 mm \textsl{BG40} filter and $10-15$ s exposures. In January and March 2012, we obtained 6.8 hr of data with 20 s exposures using GMOS \citep{Hook04} on the 8.1 m Gemini North telescope as part of the queue programs GN-2011B-Q-95 and GN-2012A-Q-29. Most of the Gemini data were taken using a Sloan-\textsl{g} filter, but we obtained nearly 2 hr of data using a Sloan-\textsl{r} filter to constrain the luminosity of the secondary WD (see Section~\ref{sec:param}). Additionally, we obtained 1.5 hr of data using 10 s exposures in March 2012 and 1.0 hr of data using 5 s exposures in April 2012 using the OSIRIS instrument \citep{Cepa00,Cepa03} through a Sloan-\textsl{g} filter and in fast photometry mode, mounted on the Gran Telescopio Canarias (GTC) 10.4 m telescope.

We bias- and flat-field correct the raw science frames using standard IRAF routines. For Argos and Agile, we perform weighted, circular, fixed-aperture photometry on the calibrated frames using the external IRAF package $\textit{ccd\_hsp}$ \citep{Kanaan02}. We divide the sky-subtracted light curves using five brighter comparison stars in the field to remove transparency variations. To remove any long-term trends caused by differential atmospheric extinction, we fit a low-order polynomial to observing runs exceeding 2 hr using the WQED software suite \citep{Thompson09}, which we also use to apply a timing correction to each observation to account for the motion of the Earth around the barycenter of the solar system \citep{Stumpff80}. We use the formalism described in \citet{EH01} to derive average point-by-point photometric errors of 1.0 mmag for GMOS and OSIRIS, and 2.8 mmag for Argos observations. We calibrate these errors using the $g_0=19.1$ mag, photometrically constant star SDSS~J065132.86+284408.4, within 20$\arcsec$ of our target. 

\begin{figure*}[t]
\centering{\includegraphics[width=0.81\textwidth]{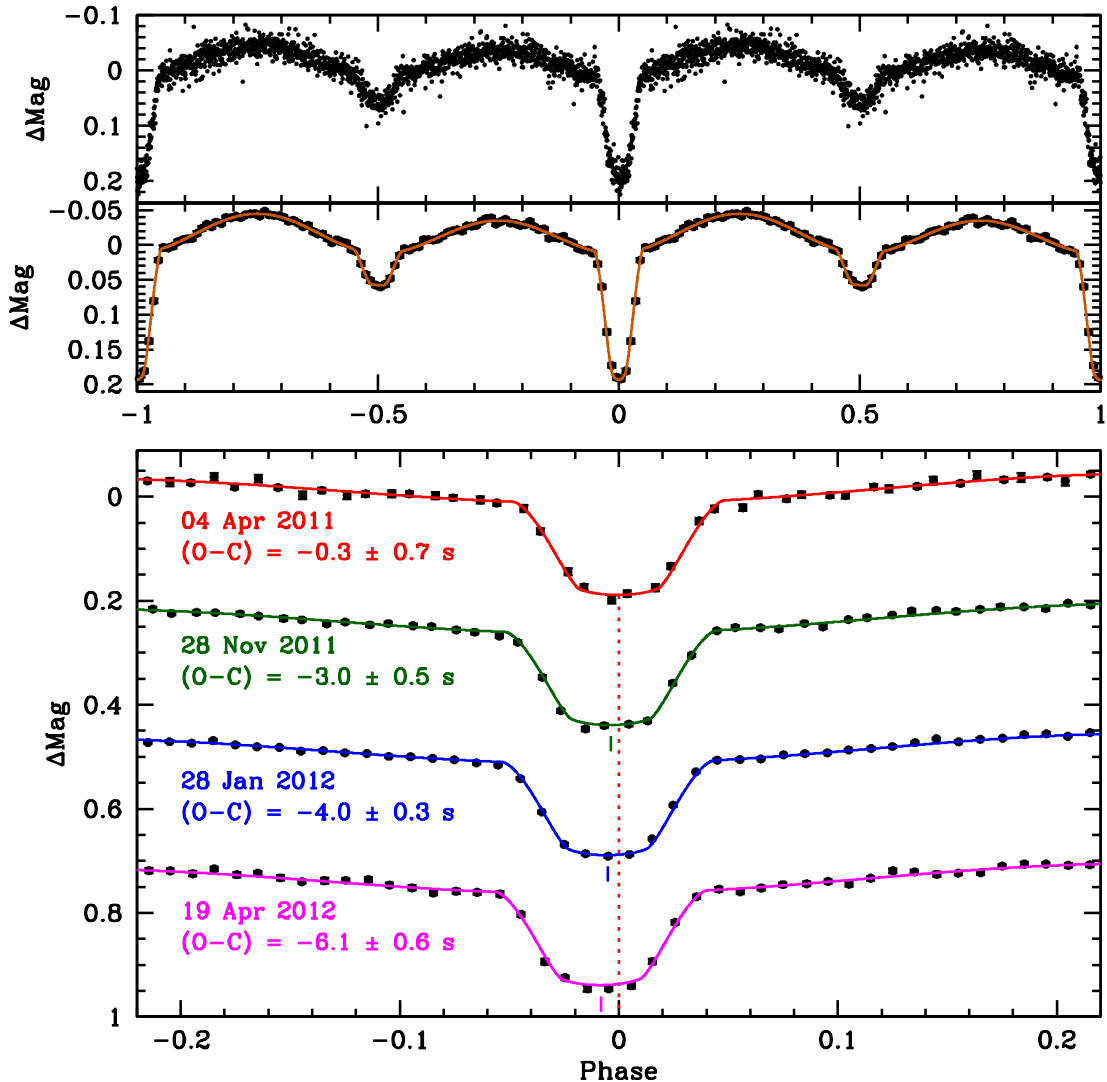}}
\caption{The top panels show high-speed photometry of J0651 from the 8.1 m Gemini North telescope and the GTC 10.4 m telescope, folded at the orbital period and duplicated for clarity. Directly below is the same data binned into 100 phase bins, with error bars, and over-plotted with our best-fit model. The bottom panel shows a portion of the folded, binned data from four different subsets, a visual representation of the \omc\ diagram in Figure~\ref{fig:J0651omc}. The decreasing orbital period is evident as the primary eclipses, shown, shift sooner. \label{fig:J0651gband}}
\end{figure*}

For the GMOS and OSIRIS data, we use DAOPHOT \citep{Stetson87} to perform aperture photometry on our target and a dozen photometrically constant SDSS point sources in our images for calibration. We use a script by \citet{Eastman10} to apply a barycentric timing correction and check it against the timings based on WQED.

\subsection{Spectroscopic Observations}

To obtain higher signal-to-noise-ratio spectroscopy and better phase coverage, we obtained additional time-series spectroscopy of J0651 at the 6.5m MMT telescope in October 2011 and April 2012. All 79 spectra were taken using the Blue Channel Spectrograph \citep{Schmidt89} with a 800 l mm$^{-1}$ grating and a 1$\arcsec$ slit. This set-up provides 2.1 \AA\ resolution and a spectral coverage from 3550---5450 \AA. The lower resolution compared to the discovery observations presented by \citet{BrownJ0651} enabled us to decrease the exposure time to 90 s, providing a radial velocity curve with better phase coverage (see Figure~\ref{fig:J0651newrv}). The reduced spectra have an average radial velocity error of $\pm$31 km s$^{-1}$.



\section{Updated System Parameters}
\label{sec:param}

We compute the orbital elements using the code of \citet{Kenyon86}, which weights each velocity measurements by its associated error. However, the observed velocity amplitude is an underestimate because our 90 s exposures span 12\% of the orbital phase, and because the radial velocity curve is not linear. By integrating a sine curve at the phase covered by our exposures, we determine that the velocity amplitude correction is 2.3\%. The resultant best-fit, corrected velocity semi-amplitude is $K=616.9\pm3.5$ km s$^{-1}$. This is significantly lower than our original value, $657.3\pm2.4$ km s$^{-1}$, computed in the same way.

Given that the original radial velocity semi-amplitude depended on a dozen measurements at quadrature with 30 km s$^{-1}$ errors, the formal error was an underestimate. A Monte Carlo calculation indicates that the true uncertainty in $K$ using our original 27 spectra was actually 14 km s$^{-1}$. The same calculation with the new data yields an uncertainty in $K$ of 5 km s$^{-1}$. This implies that the companion is less massive than originally predicted by \citet{BrownJ0651}.

We refine the physical parameters of the primary\footnote[1]{Following \citep{BrownJ0651} we refer to the low-mass WD as the primary since it contributes $>$95\% of optical light.} WD using the summed spectrum, which has S/N=78 per resolution element. Fitting our new spectra with the stellar atmosphere models of \citet{TremBer09}, which include improved Stark broadening profiles with non-ideal gas effects, formally yields \teff\ $=16530\pm105$ K and \logg\ $=6.76\pm0.02$. This result is nearly identical to our original measurements \citep[\teff\ $=16400\pm300$ K and \logg\ $=6.79\pm0.04$,][]{BrownJ0651}.


Additionally, we investigate the effect of velocity smearing on the derived atmospheric parameters by analyzing the spectra obtained at quadrature ($|v|> 500$ km s$^{-1}$, when velocity smearing should be at its minimum). For these spectra at quadrature, we find that \teff\ is 500 K lower and \logg\ is 0.07 dex higher. These differences reflect our systematic error, and also indicate how the parameters of the tidally distorted primary depend on phase, but the parameters remain consistent with the higher S/N summed and phased spectrum. Thus we adopt a mean \teff\ $=16530\pm200$ K and \logg\ $=6.76\pm0.04$, implying a $0.25$ \msun\ primary \citep{Panei07}.

We use our high-quality \textsl{g}-band Gemini and GTC photometry to refine the system parameters using the light curve fitting code JKTEBOP \citep{Southworth04}. We supply the limb-darkening coefficients from WD atmosphere models appropriate for the J0651 system using $I(\mu)/I(1) = 1 - c_1(1 - \mu) - c_2(1 - \sqrt{\mu})$, where $\mu = \cos{\theta}$ (P. Bergeron 2012, private communication). These coefficients are included in Table~\ref{tab:param}; their uncertainties are negligible given our observed \teff\ and \logg\ uncertainties. Additionally, we have adopted gravity-darkening coefficients of $\beta_1 = \beta_2 = 0.36$ for both the primary and secondary, where $F \propto T_{\mathrm{eff}}^4 \propto g^{\beta}$. We expect convection to be present in both stars, and our light curve fits do poorly for $\beta = 1.0$, as expected for a purely radiative atmosphere, so adopting $\beta = 0.36$ is reasonable.

We fix the limb- and gravity-darkening coefficients and fit for the inclination and component radii, and our error estimates result from 10,000 Monte Carlo simulations, as described in \citet{Southworth05}. Gravitational lensing should minimally affect the derived radius of the primary and secondary, by roughly 0.1\% and 0.7\%, respectively \citep{Marsh01}, and have not been included in the fits. The primary radius is a volume-average; the tidal distortions make the star 3.3\% oblate.

The photometry allows us to test the ELM WD models by providing an independent estimate on the mass of the primary. To do so, we hold fixed a series of different mass ratios in our light curve fits, and in each case use the resultant secondary radius in combination with the tested mass-radius relation of \citet{Wood95} in order to back out the mass of the primary. Consistently, this method finds $M_1 = 0.26\pm0.04$ \msun, which we adopt.

Taking $M_1 = 0.26\pm0.04$ \msun\ and $K = 616.9\pm5.0$ km s$^{-1}$, the secondary mass is thus $M_2 = 0.50\pm0.04$ \msun\ for the best-fit inclination of $84.4\pm2.3$ deg. Table~\ref{tab:param} shows our final light curve results found by fixing $q=1.92$. The resulting radius of the secondary, $R_2 = 0.0142\pm0.0010$ \rsun, implies $M_2 = 0.50\pm0.04$ \msun\ \citep{Wood95}, in good agreement. Pairing the volume-averaged primary radius $R_1=0.0371\pm0.0012$ \rsun\ with the observed surface gravity yields $M_1 = 0.29\pm0.05$ \msun, somewhat larger but consistent with our adopted value, as well as with the result using the \citet{Panei07} models.

Finally, we use our Gemini \textsl{r}-band data to constrain the luminosity and temperature of the secondary. Fixing the limb- and gravity-darkening coefficients, and adopting the inclination and component radii from the \textsl{g}-band fits, we find the secondary contributes $3.7\pm0.2$\% of light in the \textsl{g}-band and $4.6\pm0.6$\% of light in the \textsl{r}-band.

Adopting $M_g=8.9\pm0.1$ mag and $M_r=9.2\pm0.1$ mag for the 0.26 \msun\ primary \citep{Panei07}, the secondary thus has $M_g=M_r=12.5\pm0.2$ mag. For a 0.5 \msun\ WD, cooling models\footnote[3]{\href{http://www.astro.umontreal.ca/~bergeron/CoolingModels}{http://www.astro.umontreal.ca/$_{\widetilde{~}}$bergeron/CoolingModels}} suggest a temperature of $8700\pm500$ K for the secondary, which corresponds to a cooling age of roughly 700 Myr \citep{HB06,KS06,Tremblay11,Bergeron11}.



\section{Detection of Orbital Period Decay}

\begin{deluxetable}{ll}
\tabletypesize{\scriptsize}
\tablecolumns{2}
\tablewidth{0.47\textwidth}
\tablecaption{System parameters. \label{tab:param}}
\tablehead{
\colhead{Parameter} & \colhead{Value} \\ \colhead{[method used to derive parameter]} & \colhead{} }
\startdata
Orbital Period [phot.]					&	$765.206543(55)$ s		\\
$K_1$ (corrected for smearing) [spec.]	&	$616.9\pm5.0$ km s$^{-1}$			\\
$\gamma_{vel}$ [spec.]					&	$-7.7\pm4.5$ km s$^{-1}$	\\
Primary \teff\ [spec.]					&	$16530\pm200$ K  		\\
Primary \logg\ [spec.]					&	$6.76\pm0.04$			\\
Primary Mass	 ($M_1$) [phot.]		        &	$0.26\pm0.04$ \msun\	\\
Primary Radius ($R_1$) [phot.]           &   $0.0371\pm0.0012$ \rsun\                \\
Inclination	($i$) [phot.]				&	$84.4\pm2.3$ degrees		\\
Mass Ratio ($q$) [phot., spec.]          &   $1.92\pm0.46$           \\ 
Secondary Mass ($M_2$) [spec.]			&	$0.50\pm0.04$ \msun\			\\
Secondary \teff\  [phot.]				&	$8700\pm500$ K  		\\
Secondary Radius ($R_2$)	[phot.]			&	$0.0142\pm0.0010$ \rsun\	\\
Limb Darkening, Primary, $g$-band 		& $c_1 = -0.106$, $c_2 = 0.730$		\\
Limb Darkening, Secondary, $g$-band 		& $c_1 = -0.128$, $c_2 = 0.898$		\\
Limb Darkening, Primary, $r$-band 		& $c_1 = -0.076$, $c_2 = 0.562$		\\
Limb Darkening, Secondary, $r$-band 		& $c_1 = -0.099$, $c_2 = 0.735$

\enddata
\end{deluxetable}

We demonstrate a secular change in the orbital period of J0651 by constructing an \omc\ diagram, where we compare the observed mid-eclipse times ($O$) to expected mid-eclipse times computed from the assumption of a fixed orbital period ($C$) for future epochs ($E=t/P$). To estimate the mid-eclipse times, we fix the best-fit model parameters from our analysis in Section~\ref{sec:param} and fit each subset of observations only for the mid-eclipse time nearest the mean time of the observations.

\begin{deluxetable}{llcr}
\tabletypesize{\scriptsize}
\tablecolumns{4}
\tablewidth{0.47\textwidth}
\tablecaption{Journal of mid-eclipse times. \label{tab:omc}}
\tablehead{
\colhead{Facility} & \colhead{Epoch} & \colhead{No. Eclipses} & \colhead{Mid-Eclipse Time} \\
\colhead{} & \colhead{($E$)} & \colhead{} & \colhead{(BJD$_{\rm{TDB}} - 2450000$)}  }
\startdata
McD		&	373		& 59		&	5655.9015854(85)				\\
McD		&	20735	& 41		&	5836.2388047(83)				\\
McD		&	23327	& 52		&	5859.1949883(81)				\\
McD		&	27251	& 146	&	5893.9480995(57)				\\
APO,McD	&	30981	& 110	&	5926.9830564(67)				\\
Gem-N	&	31207	& 5		&	5928.984606\phantom{1}(11)	\\
McD		&	34164	& 284	&	5955.1734668(37)				\\
McD		&	36291	& 80		&	5974.0113686(59)				\\
GTC		&	39171	& 7		&	5999.518212\phantom{1}(12)	\\
McD		&	39383	& 53		&	6001.3958308(75)				\\
Gem-N	&	39426	& 6		&	6001.776617\phantom{1}(14)	\\
Gem-N	&	39542	& 10		&	6002.803962\phantom{1}(10)	\\
McD		&	43317	& 89		&	6036.2375115(72)				\\
GTC		&	43446	& 5		&	6037.3800048(72)				\\
McD		&	46578	& 16		&	6065.118768\phantom{1}(26)	
\enddata
\tablecomments{While all four instruments we have used are conditioned with a GPS receiver and thus should have absolute time stamps accurate to a few ms, the mid-exposure times for our GMOS-N data all end in .2 or .7 s, suggesting a rounding error of up to 0.25 s; the uncertainties in our Gemini mid-eclipse times have thus been enlarged by 0.25 s. We may remove any potential systematic time offsets by computing the period change using only the points from McDonald Observatory, which yields $(-8.2\pm3.2) \times 10^{-12}$ s s$^{-1}$.}
\end{deluxetable}

\begin{figure*}[t]
\centering{\includegraphics[width=0.85\textwidth]{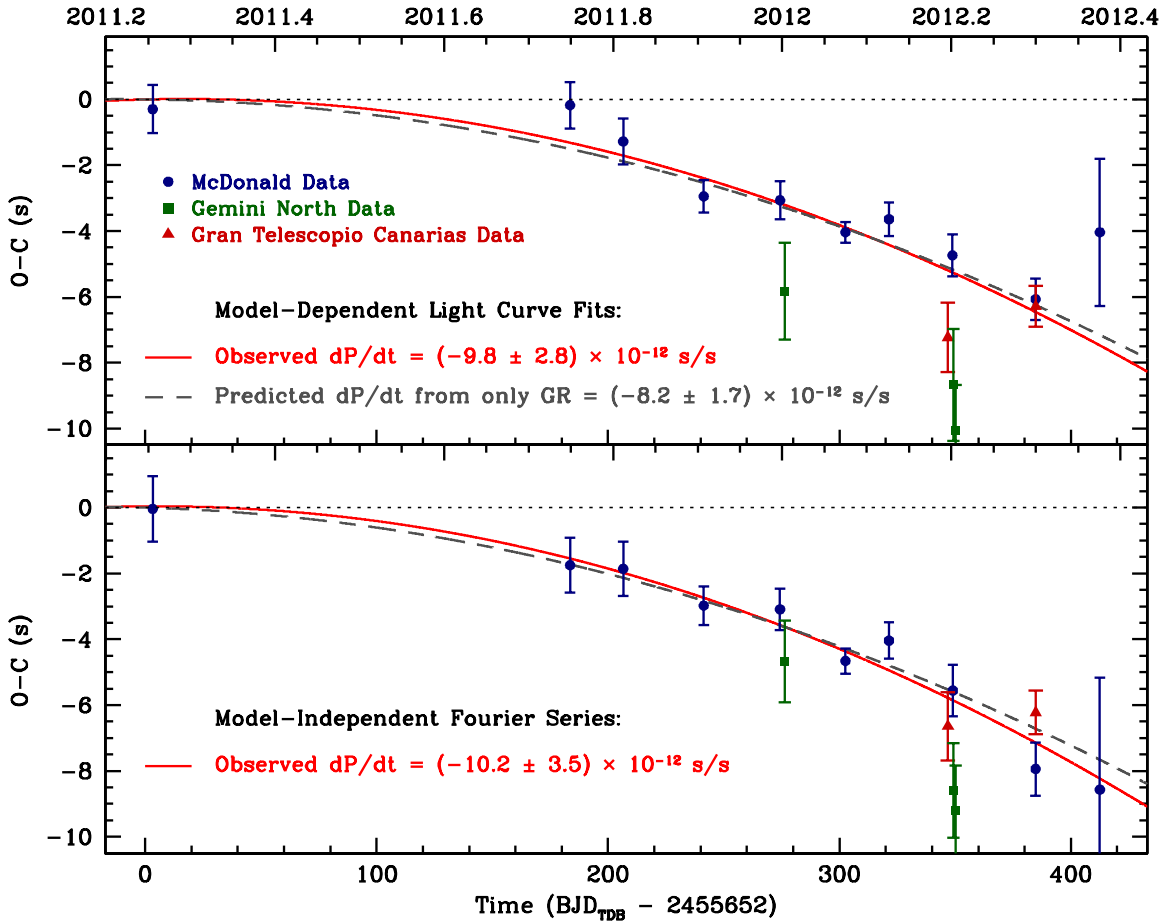}}
\caption{\omc\ diagrams of the orbital evolution in J0651 since April 2011; blue dots represent data from McDonald Observatory and APO, green squares from Gemini-North, and maroon triangles from GTC. The top panel shows the change in mid-eclipse times as determined by light curve modeling, and the best-fit parabola yields an estimate for the observed rate of orbital period change. Additionally, the bottom panel shows the results from a model-independent, linear least-squares fit using the orbital period and higher harmonics. The dotted line at $(O-C)=0$ shows the line of zero orbital decay, while the grey dashed line shows the predicted orbital decay expected solely from gravitational wave radiation. Using both methods, our early results match the GR prediction to the 1-$\sigma$ level. \label{fig:J0651omc}}
\end{figure*}

Following \citet{Kepler91}, if the orbital period is changing slowly with time, we can expand the observed mid-time of the E$^{th}$ eclipse, $t_E$, in a Taylor series around $E_0$ to arrive at the classic \omc\ equation
$$
O \; - \; C = \Delta T_{0} + \, \Delta P_0 \, E + \frac{1}{2} P_0 \dot{P} E^2 + ...
$$
where $T_0$ is the mid-time of the first eclipse, $\Delta T_{0}$ is the uncertainty in this mid-point, $P_0$ is the orbital period at the first eclipse and $\Delta P_0$ is the uncertainty in this period. Any secular change in the period, $dP/dt$, will cause a parabolic curvature in an \omc\ diagram. Currently, the acceleration in the period change, $d(dP/dt)/dt$, is negligible, and we will limit our discussion to a second-order polynomial fit.

To construct an \omc\ diagram, we must first determine $T_0$ and $P_0$. A preliminary estimate comes from a simple Fourier transform of our whole data set, which we use to create an initial \omc\ diagram. We then iteratively adjust $T_0$ and $P_0$ by the zeroth- and first-order terms from our best-fit parabola until the adjustments are smaller than the error in these terms; these errors result from the covariance matrix. Our recomputed, final \omc\ diagram uses this new ephemeris and period and is shown in Figure~\ref{fig:J0651omc}. We find:

\[T_0 = 2455652.5980910\pm0.0000084\;\rm BJD_{TDB}\]
\[P_0 = 765.206543\pm0.000055\;\rm s\]

Table~\ref{tab:omc} presents the mid-eclipse times from each subset of our observations. Each night of observing from Gemini and GTC have been given their own subset, as have each month of data from McDonald. Since $\Delta T_{0}$ and $\Delta P_{0}$ are nonzero, the zeroth- and first-order terms of the parabola indicating the predicted $dP/dt$ in Figure~\ref{fig:J0651omc} have been allowed to vary within the current constraints on these terms.

A weighted, second-order, least-squares fit to the mid-eclipse times yields a rate of period change of $(-9.8\pm2.8) \times 10^{-12}$ s s$^{-1}$ (or $-0.31\pm0.09$ ms yr$^{-1}$). This value includes our May 2012 data point, which has just 3.4 hr of data spread over four nights at a minimum airmass of 2.0. If we do not include this last point, the inferred rate of period change differs slightly, yielding $(-10.6\pm2.9) \times 10^{-12}$ s s$^{-1}$ (or $-0.33\pm0.09$ ms yr$^{-1}$). A parabola is needed to best represent the data: The best second-order fit has $\chi^2=33.0$ (12 d.o.f.), whereas the best first-order fit has $\chi^2=44.9$ (13 d.o.f.).

As a sanity check, we also construct an \omc\ diagram using a model-independent approach. Here we perform a simultaneous least-squares fit to each subset of data using a series of sine curves at the orbital period ($P_0$) up to the last harmonic before the Nyquist frequency of that subset. This effectively uses the high-amplitude ellipsoidal variations at half the orbital period as our clock, with a Fourier series of the orbital harmonics to reproduce the eclipses. A best-fit parabola to the observed minima of the ellipsoidal variations is shown in the bottom panel of Figure~\ref{fig:J0651omc} and yields $dP/dt = (-5.1\pm1.8) \times 10^{-12}$ s s$^{-1}$. In order to compare this to the mid-eclipse times at the orbital period, we must multiply this result by a factor of two, which yields $dP/dt = (-10.2\pm3.5) \times 10^{-12}$ s s$^{-1}$ (or $-0.32\pm0.11$ ms yr$^{-1}$).

This model-independent \omc\ method retains larger errors because the harmonics are inevitably truncated by the Nyquist frequency in the observations and so are only roughly capable of replicating the deep primary eclipses. Our results from this method are not orthogonal to the model-dependent light curve fitting, as both fit the eclipses and ellipsoidal variations. Therefore, both results cannot be averaged, and we emphasize the results from the model-dependent approach.

Thus, our best estimate for the rate of orbital period change in J0651 after 13 months is $(-9.8\pm2.8) \times 10^{-12}$ s s$^{-1}$ (or $-0.31\pm0.09$ ms yr$^{-1}$), a 3-$\sigma$ detection. This yields a timescale for period change, $P / \dot{P} = 2.5\pm0.8$ Myr.



\section{Discussion}

Based on the refined parameters for the J0651 system (see Section~\ref{sec:param}) and treating each WD as a point mass in a non-relativistic circular orbit, general relativity predicts an orbital period decay in this system of $(-8.2\pm1.7) \times 10^{-12}$ s s$^{-1}$ (or $-0.26\pm0.05$ ms yr$^{-1}$) \citep{LL75}. Recently, \citet{vdB12} demonstrated that the point mass approximation is valid to better than 1\% in cases such as J0651, so our uncertainty in the orbital decay from gravitational wave radiation is dominated by the uncertainty in the component masses. Additional effects that could modulate the mid-eclipse times are unlikely to explain the observed shift: For example, given the estimated distance of 1.0 kpc \citep{BrownJ0651}, we expect the proper motion to change the period by no more than $5 \times 10^{-16}$ s s$^{-1}$ \citep{Shklovskii70}.

It is evident from the high-amplitude ellipsoidal variations of the primary that strong tidal forces are also present. These tides will act as a torque to spin-up the WDs if the system is synchronized, further robbing the orbit of angular momentum and increasing the rate of orbital period decay. The degree to which this tidal torquing influences the orbital evolution depends on the effective tidal locking, which is in many ways determined by the physical structure of the ELM WD. This effect could increase the rate of period decay by at least 5\% if the system is synchronized \citep{Piro11,Benacquista11,Fuller12}. With just 13 months of monitoring, our sensitivity in the observed rate of orbital decay is not yet sufficient to detect a significant deviation from pure gravitational wave losses. However, future observations should constrain this discrepancy, providing an excellent probe of the interior of ELM WDs, in addition to the possibilities opened through asteroseismology \citep{Steinfadt10,Hermes12}.

The short period of J0651 makes it one of the loudest known sources of gravitational wave radiation, and continued monitoring of orbital decay in the system will provide strong constraints on the gravitational wave strain of J0651. This is important for future gravitational wave missions, like the evolved Laser Interferometry Space Antenna ({\em eLISA}). Critical to that success is disentangling the contributions of tidal torques on the orbital decay, an effort worthy of further photometric observations and modeling. 

\acknowledgments

We acknowledge the anonymous referee for valuable suggestions that greatly improved this manuscript. The authors are especially grateful to those who contributed to these observations: John W. Kuehne, K. I. Winget, E. L. Robinson, Paul A. Mason, and Samuel T. Harrold. C.A.P. is thankful to Alessandro Ederoclite for suggesting to observe J0651 with GTC, and to Sebasti\'an Hidalgo for help with photometric reductions. J.J.H., M.H.M. and D.E.W. acknowledge the support of the NSF under grant AST-0909107 and the Norman Hackerman Advanced Research Program under grant 003658-0252-2009. A.S.M. acknowledges observing support from NSF grant AST-1008734. 

{\it Facilities:} \facility{McDonald 2.1m (Argos), MMT (Blue Channel Spectrograph), Gemini North (GMOS), GTC (OSIRIS), APO (Agile)}

\end{document}